\documentclass[a4paper]{revtex4}
\usepackage{graphicx}
\usepackage{fancyhdr}
\usepackage{amsmath}
\usepackage{color}

\begin{document}

\title{From the Higgs to the top: Couplings and Rare Decays }

%

\author{J. Lorenzo Diaz-Cruz}
\affiliation{Facultad de Ciencias F\'isico-Matem\'aticas\\
Benem\'erita Universidad Aut\'onoma de Puebla, C.P. 72570, 
Puebla, Pue., Mexico.}

\begin{abstract}
Within the Standard Model the Higgs couplings to fermions and gauge bosons, 
as function of the particle mass, are predicted to lay on  a single line. However, new 
patterns appear in multi-Higgs models that employ the Froggart-Nielsen mechanism, where 
the diagonal couplings  could lay on different lines  and flavor-violating Higgs couplings 
could appear too.  These aspects are studied for  a specific model with 3+1 Higgs doublets 
and a singlet FN field.  Constraints on the model are derived from the Higgs search at LHC, 
and some remarks are presented on their implications for the rare top and Higgs decay, 
$t\to ch$ and $h\to \tau \mu$.  

\end{abstract}

\maketitle

\thispagestyle{fancy}


\section{Introduction}

The discovery of a Higgs-like particle with $m_h=125-126$ GeV at the 
LHC \cite{higgs-atlas:2012gk,higgs-cms:2012gu}, has verified 
the mechanism of Electro-Weak symmetry breaking of the 
Standard Model (SM) \cite{Gunion:1989we}.  
Current measurements of the spin, parity, and couplings of the Higgs, seem  consistent with 
the SM.  On the other hand, several scenarios for Physics Beyond the SM (PBSM) have been proposed 
to address some of its open problems, such as hierarchy, flavor, unification, 
etc \cite{Pomarol:2012sb,Martin:1997ns}. So far, the LHC  bounds on its 
scale ($\Lambda$)  are entering into the multi-TeV range, which could mean $\Lambda >> v=246$ GeV. 
A more solid conclusion needs
to wait for the next LHC run, with higher energy and luminosity.
Many papers have been devoted to study the LHC implications for the Higgs couplings, for instance in
\cite{Espinosa:2012ir,Giardino:2013bma}. 
The couplings of the Higgs particle to a pair of massive gauge bosons or fermions, 
are proportional  to the particle mass.  However,  the LHC has tested only a few of these couplings,
i.e. the ones with the heaviest SM fermions and $W,Z$. Non-standard Higgs couplings, 
including the flavor violating (FV) ones, are 
predicted in many models of physics beyond the SM, for instance 
in the general multi-Higgs models  \cite{Branco:2011iw, DiazCruz:2004tr}
and SUSY \cite{DiazCruz:2002er}.

 Many  ideas have been proposed to adress the flavor problem \cite{Isidori:2010kg}, for instance: 
Textures and GUT-inspired relations, flavor symmetries and radiative generation. 
Within the flavor symmetry approach,  the Froggart-Nielsen mechanism assumes that above some scale $M_F$, 
such symmetry forbids the  appearence of the Yukawa couplings; 
SM fermions are charged under this symmetry (which could be of Abelian type $U(1)_F$).
However, the Yukawa matrices can arise  through  non-renormalizable operators. 
The Higgs spectrum of these models could include a  light and heavy Higgs boson. 
In these models the diagonal flavor conserving (FC) couplings of the light SM-like Higgs boson 
could deviate from SM, while flavor violating (FV) couplings could be induced at small rates too. 
Within the SM,  the FC Higgs couplings to fermions and gauge bosons, 
as function of the particle mass, lay on a single straight line. 
However, in multi-Higgs  models, they could lay on distinct lines.
 As illustrated in Figure 1, the scalar spectrum could include:

i) A lightest state that should be identified with the SM-like Higgs boson. One expects, 
the appearence of small deviations from SM predictions for the  diagonal Higgs-fermion couplings 
and suppressed FV couplings. 

ii)  States with Flavon-dominated composition, which could provide the more radical 
signature of the models under consideration.
 The observation of these signals depends on the flavon scale, and could at the reach of
 the LHC reach if such scale were about $O(1)$ TeV. 
 
iii)  Heavy Higgs bosons  which could have large mixing with flavons, and thus 
deviate significantly from SM  expectations for  FV couplings, 
that could also be searched at LHC.

\begin{figure}[ht]
\centering
\includegraphics[width=1.6in]{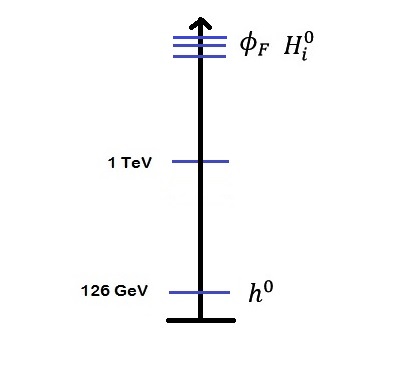}
\includegraphics[width=1.6in]{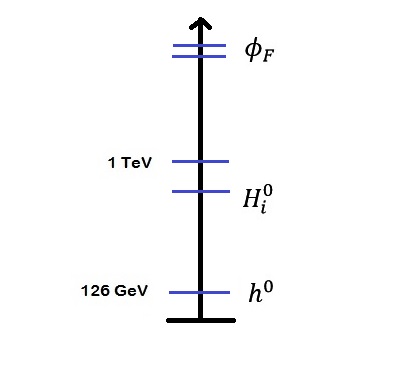}
\includegraphics[width=1.9in]{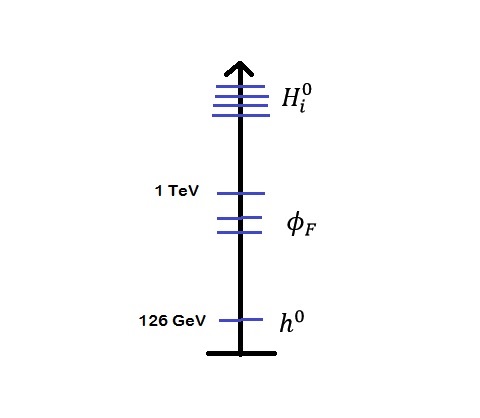}
\caption{ The Higgs and flavon spectrum in multi-Higgs models.}
\label{ }
\end{figure}

\section{Higgs couplings within a 3+1 Higgs model}

Thus, we shall consider a 3+1-Higgs doublet model, denoted as $\Phi_0,\Phi_1, \Phi_2, \Phi_3$.
The $\Phi_1$ gives masses to up-type quarks, while 
$\Phi_2$ and $\Phi_3$ give masses to d-type quarks
and leptons, respectively.  We shall impose a discrete symmetry in such a way that  one doublet ($\Phi_0$)
is of the inert-type, and therefore contains a Dark matter condidate \cite{Diaz-Cruz:2014pla}. 
Furthermore, we shall also include Froggart-Nielsen scalar field (SM singlet $S$). 
The possibility of having light flavon fields was studied in ref. 
\cite{Dorsner:2002wi}, and more recently in \cite{Tsumura:2009yf, Berger:2014gga}.
The Yukawa lagrangian is given by:
\begin{equation}
 {\cal{L}}_Y =    \rho^u_{ij}  ( \frac{ S }{\Lambda_F} )^{n_{ij}} \bar{Q}_i d_j  \tilde{\Phi}_1
                + \rho^d_{ij}  (\frac{ S }{\Lambda_F})^{p_{ij}} \bar{Q}_i u_j \Phi_2  
                + \rho^l_{ij}  (\frac{ S }{\Lambda_F})^{q_{ij}} \bar{L}_i l_j \Phi_3  +h.c.
\end{equation}
where $n,p,q$ denote the charges of each fermion type under an Abelian flavor 
symmetry, which will help to explain the fermion mass hierarchy. 
The flavon field $S$ is assumed to have flavor charge equal to -1, such that 
${\cal{L}}_{Y}$ is $U(1)_F$-invariant.
Then,  Yukawa couplings arise after the spontaneous breaking  of the flavor symmetry,
i.e. $\lambda_x = (\frac{<S>}{\Lambda_F})^{n_x}$, where $<S>$ denotes the
flavon vacuum expectation  value, while $M_F$ denotes the heavy mass 
scale, which represents the mass of heavy fields that  
transmit such symmetry breaking to the quarks and leptons.

The Higgs and Flavon fields are written in terms of mass eigenstates, through the
rotation  $O^T$ $(4\times 4)$: 
\begin{eqnarray}
Re\Phi^0_i &=&  O^T_{i1} h^0_1 + O^T_{i2} H^0_2 + O^T_{i3} H^0_3 + O^T_{i4} H^0_F \nonumber \\
Re S       &=&  O^T_{41} h^0_1 + O^T_{42} H^0_2 + O^T_{43} H^0_3 + O^T_{44} H^0_F
\end{eqnarray}
Furthermore, as the vev's must satisfy: $v^2_1+ v^2_2+v^2_3 = v^2$, with $v=246$ GeV, 
we find convenient to use spherical coordinates to express each vev ($v_i$) in terms of the
total vev $v$ and the angles $\beta_1$ and $\beta_2$, as shown in figure 2, namely:
$v_1 = v \cos\beta_1$, $v_2 = v \sin \beta_1 \cos \beta_2$ and 
$v_3 = v \sin \beta_1 \sin \beta_2$. 

\begin{figure}[ht]
\centering
\includegraphics[width=2.2 in]{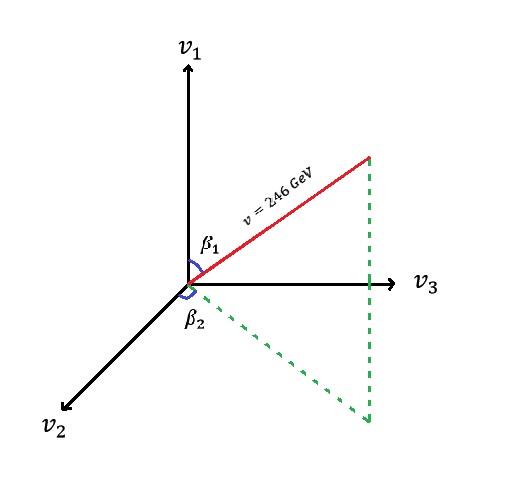}
\caption{ The Higgs vevs in spherical coordinates.}
\label{ }
\end{figure}

Thus, for the lightest Higgs state ($h^0_1=h^0$, one gets finally
the following interaction lagrangian for the Higgs-fermion couplings,
\begin{equation}
 {\cal{L}}_Y =  [ \frac{\eta^u } {v} \bar{U} M_u U  + \frac{\eta^d  }{v} \bar{D} M_d  D  
             + \frac{\eta^l }{v}  \bar{L} M_l L  
            +  \kappa^u \bar{U}_i \tilde{Z}^u U_j  + \kappa^d \bar{D}_i \tilde{Z}^dD_j 
                + \kappa^l   \bar{L}_i\tilde{Z}^l  L_j ] h^0  
\end{equation}
where: $\eta^u = O^T_{11}/\cos\beta_1$, $\eta^d=O^T_{21}/\sin \beta_1 \cos \beta_2 $, 
$\eta^l=O^T_{31}/ \sin \beta_1 \sin \beta_2$, describe the strength of the flavor-diagonal
Higgs couplings. While the FV Higgs couplings are described by the parameters:
$\kappa^u = \frac{v}{u} O^T_{41} \cos\beta_1 $, \, 
$\kappa^d = \frac{v}{u} O^T_{41} \sin\beta_1 \cos \beta_2 $, \, 
$\kappa^l = \frac{v}{u} O^T_{41} \cos\beta_1 \sin \beta_2 $.  

Besides the Yukawa couplings, we also need to specify the Higgs couplings with vector bosons,
which is written as $g_{hVV} =  \chi_V g^{sm}_{hVV} $, with the factor $\chi_V$ 
given as:
\begin{eqnarray}
 \chi_V &=&  \frac{v_1}{v}O^T_{11} + \frac{v_2}{v}O^T_{21}+ \frac{v_3}{v}O^T_{31} \nonumber \\
     &=&  \cos\beta_1 \, O^T_{11} + \sin\beta_1 \cos\beta_2 \, O^T_{21}+ \sin\beta_1 \sin\beta_2 \, O^T_{31}
\end{eqnarray}

It is interesting to note that the coupling $\chi_V$ can be  written in terms of the  FC fermionic couplings,
which can be seen as a type of sum rule, i.e.
\begin{equation}
 \chi_V = \cos^2\beta_1 \, \eta^u + \sin^2\beta_1 \cos^2\beta_2 \, \eta^d + \sin^2\beta_1 \sin^2\beta_2 \, \eta^l
\end{equation}

Moreover, since the Higgs couplings to first generation quarks and leptons is highly suppressed,
in order to study the FV Higgs coupling, which depends on the matrices $\tilde{Z}^f$, 
we shall consider the 2-3 family sub-system. Namely, for up quarks the $Z$-matrix
(in mass eigenstate basis),  is given by:

\begin{equation}
\tilde{Z}^{u}=
 \left( \begin{array}{cc}
  Y^u_{22}      &    Y^u_{23}   \\
  Y^u_{23}      &    2 s_u Y^u_{23}   
\end{array} \right)
\end{equation}

and similarly for d-quarks and leptons. We find a relation among
the parameters, such that we can express the $\rho^{u,d}_{ij}$'s in terms of the ratios of masses 
and the CKM angle $V_{cb} \simeq s_{23}$. Namely, we define:  $r_u= m_c/m_t$, $r_d= m_s/m_b$, 
and  $r^u_1= Y^u_{22}/Y^u_{33}$, $r^u_2= Y^u_{23}/Y^u_{33}$. 
Similarly:
$r^d_1= Y^d_{22}/Y^d_{33}$, $r^d_2= Y^d_{23}/Y^d_{33}$.
Within this approximation we have: 
$\tilde{Y}^f_{33} \simeq Y^f_{33}$ for $f=u,d$.
Then,
$r^f_1= r_f+ r^f_2$, and the ratios of Yukawas must satisfy the following relation:
\begin{equation}
r^u_2= r^d_2 \frac{1+r_d}{1+r_u} - \frac{s_{23}}{1+r_u} 
\end{equation}

Thus, in order to study the predictions of our model, we need to specify the vevs $v_i$ and the 
rotation matrix for Higgs particles ($O_{ij}$). For  the 2HDM
(see for instance \cite{Ferreira:2014naa}), LHC Higgs data favors both 
decoupling and  alignment solutions, namely both $\tan\beta \simeq 1$ and $\tan\beta >> 1$ 
are acceptable solutions. 
Thus, for the vev's we leave $\beta_1$ as free parameter, then explore the following cases:
\begin{itemize}
 \item (VEV1) We can take  first $v_2=v_3$, which in  spherical coordinates, means:
        $\beta_2=\frac{\pi}{4}$,
 \item (VEV2) We also consider vevs with $v_2<v_3$, for which we take: $\beta_2=\frac{\pi}{3}$,
 \item (VEV2) We also consider vevs with $v_2>v_3$, for which we take: $\beta_2=\frac{\pi}{6}$,
 \end{itemize}
 Then, for the rotation matrix $O$ of real components of scalar fields, we can identify several 
 interesting scenario where the 126-Higgs is   lighter than the heavy Higgs particles and
        the flavons,  i.e.  $m_h <  m_{H_i} \simeq m_{H_F}$, which have masses of order TeV.
 Here, we shall consider a special sub-case, namely we shall assume that 
 $O^T_{11} > O^T_{i1}$, and will use the orthogonality relation for 
 the rotation matrix $O$, in order to relate the parameters, namely:
 $ (O^T_{11})^2 +  (O^T_{21})^2+ (O^T_{31})^2+ (O^T_{41})^2 = 1 $.  
 Furthermore, assuming $O^T_{i1}\simeq O^T_{j1}$ (for $i\neq j$) one has
 $ O^T_{j1} = \sqrt{ \frac{1 -  (O^T_{11})^2}{3}} $.

\section{Flavor conserving Higgs couplings at LHC}

LHC data on Higgs boson has been used to derive bounds on the Higgs couplings,
i.e. deviation from the SM, which are defined as:
$g_{hXX}= g^{sm}_{hXX} (1 + \epsilon_X)$,
where $\eta^X= 1 + \epsilon_X$. A complete analysis is done in ref. \cite{Giardino:2013bma};
for fermions, they obtain the allowed values:
$\epsilon_t= -0.21\pm 0.23$, $\epsilon_b= -0.19\pm 0.3$, $\epsilon_{\tau}= 0 \pm 0.18$;
while for W (Z) bosons they find: $\epsilon_W= -0.15\pm 0.14$, $\epsilon_Z= -0.01\pm 0.13$.

An extensive analysis of parameters satisfying these bounds will be presented elsewhere, 
with detailed numerical scans; here we shall  pick a few specific points in parameter space, 
which satisfy the LHC bounds, and will help us to  understand qualitatively the beahivour 
of the model. These poins will also be used in the next section in our analys of FCNC top decays. 
Thus, we show in figure 3-6 the predictions for each of these parameters, as function of the 
angle $\beta_1$, for the case with  $\beta_2= \frac{\pi}{3}, \, \frac{\pi}{4}, \,\frac{\pi}{6}$, 
and for $O_{11} = 0.5, \, 0.75, \, 0.9$. We can see that it is possible to satisfy these bounds for all the
$\epsilon$'s.

\begin{figure}[ht]
\centering
\includegraphics[width=2.5in]{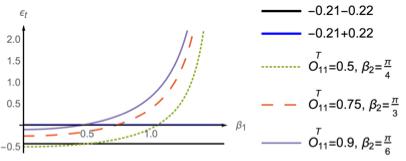}
\includegraphics[width=2.5in]{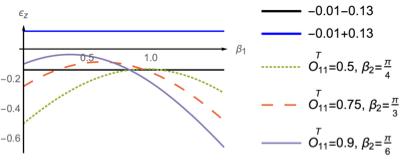}
\caption{
Higgs coupling factors $\epsilon_t$ and $\epsilon_Z$  for the parameters of set 1.
The horizontal lines are the experimental limits on these factors.}
\label{ }
\end{figure}

\begin{figure}[ht]
\centering
\includegraphics[width=2.5in]{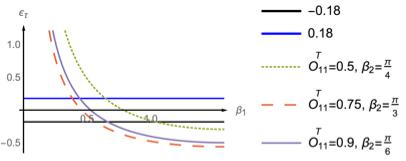}
\includegraphics[width=2.5in]{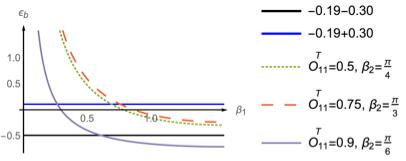}
\caption{
Higgs coupling factors $\epsilon_{\tau}$ and $\epsilon_b$ for the parameters of set 1.
The horizontal lines are the experimental limits on these factors.}
\label{ }
\end{figure}

One specific point, in agreement with all data, is :  $\beta_1=0.5$ with $O_{11} =0.9$ and $\beta_2=\frac{\pi}{6}$. 
For these values we have:
$\eta^u=1.03$, $\eta^d=0.6$ and $\eta^l=1.04$ and $\chi_v=0.96$. 
This shows that $h$ behaves very much SM-like, except for the coupling with d-type quarks.
Then, using these values we can plot the Higgs-fermion coupling as function of the mass,
as shown  in figure 2. We can see that the couplings for each fermion type lay on 
different lines, which could be distinguished from the SM (Black line). 
Future meassurements of these couplings at LHC Run2, ILC or FCC will help us in 
order to discriminate betwen our model and the SM.

\begin{figure}[ht]
\centering
\includegraphics[width=2.5in]{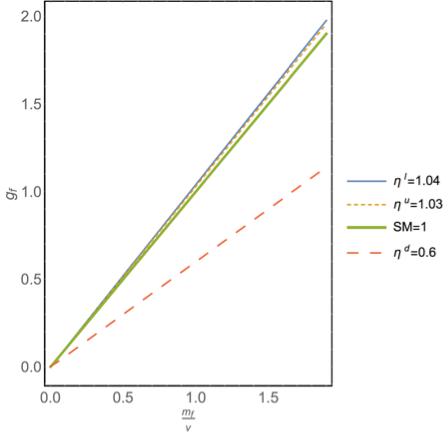}
\caption{ The Higgs-fermion coupling factors as function of mass, for parameters 
defined in the text. SM case (Black), up-type quarks (red), d-type quarks (blue), charged leptons (green).}
\label{ }
\end{figure}

We also find that the corrections contained
in the factors $\kappa^f \tilde{Z}^f$, will not change significantly the
above discussion for the top quark-Higgs couplings. However,
the Higgs coupling with the lighter fermions ($b\bar{b}, c\bar{c}, \tau^+\tau^-$),
could be measured at next-linear collider (NLC) with a precision 
of a few percent, and it will be possible to test these effects. 
The corrections to the coupling $h\bar{b}b$, could
modify the dominant decay of the light Higgs,
as well as the associated production of the Higgs with 
b-quark pairs \cite{DiazCruz:1998qc}.

We shall consider the following sample values: $r^2_d= 0.05, 0.1, 0.3$, and also
assume: $\cos\beta_1\simeq 1$, then table 1 shows the values
of the entries for the $\tilde{Z}^u$ matrix for the 2nd-3rd family case. We choose to focus 
on the up-quark sector, because we want to get an estimate for the most relevant
predictions of the model, which we believe is related with the top quark physics,
and in particular for the decay $t\to c+h$.
For the specific point in parameter space, presented in previous section:
 $\beta_1=0.5$ with $O_{11} =0.9$ and $\beta_2=\frac{\pi}{6}$. 
which is in agreement with LHC data,
we obtain the following value $\kappa^u=0.23 \frac{v}{u}$.

\section{The FCNC decay $t\to ch$ }

 The top radiative decay $t\to c+\gamma$ was first 
calculated in  \cite{LDCtopfv}, followed by ref. \cite{Eilam:1991} which presented the complete
calculations of the FCNC modes $t \to cX$ ($X=\gamma,g,Z,h$); for corrections to SM results for
$t\to ch$ and SUSY results see also: 
\cite{Mele:1998,DiazCruz:2001gf}. The 3-body FCNC decays modes $t \to cW^+W^- (ZZ,\gamma\gamma)$, were 
presented in ref. \cite{Jenkins:1996zd,DiazCruz:1999ab}, while the mode 
$t \to c\ell^-\ell^+$ was discussed recently in \cite{Frank:2006, DiazCruz:2012xa}. 
The 4-body decay $t\to bW\ell^-\ell^+ $ was also studied recently \cite{Quintero:2014lqa}.
Here we shall focus on the mode $t\to ch$, which can reach large BR's. The decay with for 
$t\to ch$ within our model is given by:
\begin{equation}
 \Gamma (t\to ch) = \frac{m_t}{6 \pi}  |\kappa^u \tilde{Z}_{23}|^2
\end{equation}
 Using the value $\Gamma(t\to b+W) \simeq 1.5$ GeV, we obtain:
$BR(t\to ch)=0.58 |\kappa^u \tilde{Z}_{23}|^2$. 
For $v/u=0.25$, and  one finds
that the B.R. could reach a value $BR \simeq 1.5\times 10^{-4}$, 
which could be tested at LHC \cite{Greljo:2014dka}.
Values of BR for other choices of parameters are shown
in table 1. 

\begin{center}
\begin{table}
\begin{center}
\begin{tabular}{| c| c | c | c |}
\hline
 Scenario  & u[TeV] & $\kappa^u \times \tilde{Z}_{23}$ & $B.R.(t\to ch)$  \\ 
 \hline
   X1     & 0.5  & $1.2 \times 10^{-4}$    & $8.6 \times 10^{-9}$  \\ 
 \hline
   X2     & 1    & $6.1 \times 10^{-5}$    & $2.2 \times 10^{-9}$  \\ 
 \hline
   X3     & 10   & $6.1 \times 10^{-6}$    & $2.2 \times 10^{-11}$  \\ 
 \hline
   Y1     & 0.5  & $6.9 \times 10^{-3}$    & $2.7 \times 10^{-5}$  \\ 
 \hline
   Y2     & 1    & $3.4 \times 10^{-3}$    & $6.8 \times 10^{-6}$  \\ 
 \hline
   Y3      & 10  & $3.4 \times 10^{-4}$    & $6.8 \times 10^{-8}$  \\ 
 \hline
   Z1     & 0.5  & $2.9 \times 10^{-2}$    & $4.8 \times 10^{-4}$  \\ 
 \hline
   Z2     & 1    & $1.4 \times 10^{-2}$    & $1.2 \times 10^{-4}$  \\ 
 \hline
   Z3      & 10  & $1.4 \times 10^{-3}$    & $1.2 \times 10^{-6}$  \\ 
 \hline  
\hline
\end{tabular}
\end{center}
\caption{ The factor $\kappa^u \times \tilde{Z}^u_{23}$ and Branching ratios for $t\to ch$}
\label{Tchbr}
\end{table}
\end{center}

\section{The LFV decay $h\to \tau \mu$ }
 
Another interesting probe of FV Higgs couplings is provided by the decay $h \to \tau \mu$, 
which was initially studied in refs. \cite{Pilaftsis:1992st, DiazCruz:1999xe}. Subsequent studies on 
detectability of the signal appeared soon after \cite{Han:2000jz, Assamagan:2002kf, Kanemura:2005hr}. 
Precise loop calculations with massive neutrinos, SUSY and other models appeared in 
\cite{Arganda:2004bz,DiazCruz:2002er,Brignole:2004ah,DiazCruz:2008ry}.
The recent search for this decay at LHC \cite{Khachatryan:2015kon}, have resulted in a bound
for the corresponding branching ratio of order $B.r.(h\to \tau\mu) < 1.51 \times 10^{-2}$  at 95\% c.l..
Furthermore, given that the best fit to the data gives $B.r.(h\to \tau\mu) = 0.84^{+0.39}_{-0.37} \times 10^{-2}$, 
many more papers have appeared recently, trying to explain this result \cite{Vicente:2015cka}. 
The search for this LFV Higgs decay could be one great opportunity to find new physics at the LHC RunII.

\section{Conclusions and outlook}

We have studied the Higgs couplings, within a model with 3+1 Higgs doublets, 
where the masses for each fermion type,
arise from a different Higgs doublet. This model also includes mixing of the Higgs 
doublets with a Flavon field, which generates the Yukawa hierarchies and induces 
Flavor-violating Higgs couplings at acceptable rates.
Constraints on these couplings, derived from Higgs search at LHC, and their implications
for FCNC top decay $t\to ch$, were discussed too. We find that this mode could reach
a BR of order $10^{-4}$, which could be studied at LHC. In the down-quark and lepton sectors, 
there are also interesting aspects to study in the future, such as the rates for rare b-decays. 
or  the decay $h \to \tau\mu$, which can be induced at rates 
that could be detected at future colliders.


\begin{acknowledgments}
Many thanks to S. Kanemura and all the organizers of HPNP15 for making my visit to Toyama a great experience. 
Support from CONACYT-SNI (Mexico) and VIEP(BUAP), and discussions with
E. Diaz, Olga Felix, B. Larios and R. Noriega-Papaqui are acknowledged.
\end{acknowledgments}


\begin{thebibliography}{99}

\bibitem{higgs-atlas:2012gk}
  G.~Aad {\it et al.}  [ATLAS Collaboration],
  Phys.\ Lett.\ B {\bf 716}, 1 (2012)
  [arXiv:1207.7214 [hep-ex]].


\bibitem{higgs-cms:2012gu}
  S.~Chatrchyan {\it et al.}  [CMS Collaboration],
  Phys.\ Lett.\ B {\bf 716}, 30 (2012)
  [arXiv:1207.7235 [hep-ex]].

\bibitem{Gunion:1989we}
  J.~F.~Gunion, H.~E.~Haber, G.~L.~Kane and S.~Dawson,
  Front.\ Phys.\  {\bf 80}, 1 (2000).

\bibitem{Pomarol:2012sb}
  A.~Pomarol,
  CERN Yellow Report CERN-2012-001, 115-151
  [arXiv:1202.1391 [hep-ph]].


\bibitem{Martin:1997ns}
  S.~P.~Martin,
  In *Kane, G.L. (ed.): Perspectives on supersymmetry II* 1-153
  [hep-ph/9709356].
  
  

\bibitem{Espinosa:2012ir} 
  J.~R.~Espinosa, C.~Grojean, M.~Muhlleitner and M.~Trott,
  JHEP {\bf 1205}, 097 (2012)
  [arXiv:1202.3697 [hep-ph]].
  
\bibitem{Giardino:2013bma} 
  P.~P.~Giardino, K.~Kannike, I.~Masina, M.~Raidal and A.~Strumia,
  arXiv:1303.3570 [hep-ph].
  
  

\bibitem{Branco:2011iw}
  G.~C.~Branco, P.~M.~Ferreira, L.~Lavoura, M.~N.~Rebelo, M.~Sher and J.~P.~Silva,
  Phys.\ Rept.\  {\bf 516}, 1 (2012)
  [arXiv:1106.0034 [hep-ph]].


  
\bibitem{DiazCruz:2004tr}
   J.~L.~Diaz-Cruz, R.~Noriega-Papaqui and A.~Rosado,
  Phys.\ Rev.\ D {\bf 69}, 095002 (2004)
  [hep-ph/0401194].

\bibitem{Isidori:2010kg} 
  G.~Isidori, Y.~Nir and G.~Perez,
  Ann.\ Rev.\ Nucl.\ Part.\ Sci.\  {\bf 60}, 355 (2010)
  [arXiv:1002.0900 [hep-ph]].
  

\bibitem{Dorsner:2002wi} 
  I.~Dorsner and S.~M.~Barr,
  Phys.\ Rev.\ D {\bf 65}, 095004 (2002)
  [hep-ph/0201207].
    

\bibitem{Diaz-Cruz:2014pla} 
  J.~L.~Diaz-Cruz,
  arXiv:1405.0990 [hep-ph].

    
\bibitem{Tsumura:2009yf} 
  K.~Tsumura and L.~Velasco-Sevilla,
  Phys.\ Rev.\ D {\bf 81}, 036012 (2010)
  [arXiv:0911.2149 [hep-ph]].
   

\bibitem{Berger:2014gga} 
  E.~L.~Berger, S.~B.~Giddings, H.~Wang and H.~Zhang,
  Phys.\ Rev.\ D {\bf 90}, no. 7, 076004 (2014)
  [arXiv:1406.6054 [hep-ph]].
   

\bibitem{Ferreira:2014naa} 
  P.~M.~Ferreira, J.~F.~Gunion, H.~E.~Haber and R.~Santos,
  arXiv:1403.4736 [hep-ph].

\bibitem{DiazCruz:1998qc} 
  J.~L.~Diaz-Cruz, H.~J.~He, T.~M.~P.~Tait and C.~P.~Yuan,
  Phys.\ Rev.\ Lett.\  {\bf 80}, 4641 (1998)
  [hep-ph/9802294].

 
\bibitem{LDCtopfv}
  J.~L.~D\'iaz-Cruz, R.~Martinez, M.~A.~Perez and A.~Rosado,
  Phys.\ Rev.\ D {\bf 41}, 891 (1990).
  
\bibitem{Eilam:1991}
G. Eilam, J. L. Hewett and A. Soni, Phys. Rev. D \textbf{44}, 1473 (1991) [Erratum-ibid. D
\textbf{59}, 039901 (1999)].

\bibitem{Mele:1998}
B. Mele, S. Petrarca, and A. Soddu, Phys. Lett. B \textbf{435}, 401 (1998)
{[hep-ph/9805498]}. 



\bibitem{DiazCruz:2001gf} 
  J.~L.~Diaz-Cruz, H.~-J.~He and C.~P.~Yuan,
  Phys.\ Lett.\ B {\bf 530}, 179 (2002)
  [hep-ph/0103178].
  


\bibitem{Jenkins:1996zd} 
  E.~E.~Jenkins,
  Phys.\ Rev.\ D {\bf 56}, 458 (1997)
  [hep-ph/9612211].


\bibitem{DiazCruz:1999ab} 
  J.~L.~Diaz-Cruz, M.~A.~Perez, G.~Tavares-Velasco and J.~J.~Toscano,
  Phys.\ Rev.\ D {\bf 60}, 115014 (1999)
  [hep-ph/9903299].
  

\bibitem{Frank:2006}
M. Frank and I. Turan, Phys. Rev. D \textbf{74}, 073014 (2006)
{[hep-ph/0609069]};

\bibitem{DiazCruz:2012xa} 
  J.~L.~Diaz-Cruz, A.~Diaz-Furlong, R.~Gaitan-Lozano and J.~H.~Montes de Oca Y.,
  Eur.\ Phys.\ J.\ C {\bf 72}, 2119 (2012)
  [arXiv:1203.6889 [hep-ph]].


\bibitem{Quintero:2014lqa} 
  N.~Quintero, J.~L.~Diaz-Cruz and G.~Lopez Castro,
  arXiv:1403.3044 [hep-ph].


\bibitem{Greljo:2014dka} 
  A.~Greljo, J.~F.~Kamenik and J.~Kopp,
  arXiv:1404.1278 [hep-ph].
    
    
\bibitem{Pilaftsis:1992st} 
  A.~Pilaftsis,
  Phys.\ Lett.\ B {\bf 285}, 68 (1992).

\bibitem{DiazCruz:1999xe} 
  J.~L.~Diaz-Cruz and J.~J.~Toscano,
  Phys.\ Rev.\ D {\bf 62}, 116005 (2000)
  [hep-ph/9910233].
  
\bibitem{Han:2000jz} 
  T.~Han and D.~Marfatia,
  Phys.\ Rev.\ Lett.\  {\bf 86}, 1442 (2001)
  [hep-ph/0008141].
  
\bibitem{Assamagan:2002kf} 
  K.~A.~Assamagan, A.~Deandrea and P.~A.~Delsart,
  Phys.\ Rev.\ D {\bf 67}, 035001 (2003)
  [hep-ph/0207302].

\bibitem{Kanemura:2005hr} 
  S.~Kanemura, T.~Ota and K.~Tsumura,
  Phys.\ Rev.\ D {\bf 73}, 016006 (2006)
  [hep-ph/0505191].
  
    
  
\bibitem{Arganda:2004bz} 
  E.~Arganda, A.~M.~Curiel, M.~J.~Herrero and D.~Temes,
  Phys.\ Rev.\ D {\bf 71}, 035011 (2005)
  [hep-ph/0407302].

\bibitem{DiazCruz:2002er}
  J.~L.~Diaz-Cruz,
  JHEP {\bf 0305}, 036 (2003)
  [hep-ph/0207030];
  
  
\bibitem{Brignole:2004ah} 
  A.~Brignole and A.~Rossi,
  Nucl.\ Phys.\ B {\bf 701}, 3 (2004)
  [hep-ph/0404211].
  
  
\bibitem{DiazCruz:2008ry} 
  J.~L.~Diaz-Cruz, D.~K.~Ghosh and S.~Moretti,
  Phys.\ Lett.\ B {\bf 679}, 376 (2009)
  [arXiv:0809.5158 [hep-ph]].

\bibitem{Cotti:2002zq} 
  U.~Cotti et al,
  Phys.\ Rev.\ D {\bf 66}, 015004 (2002)
  [hep-ph/0205170].
      
  
\bibitem{Khachatryan:2015kon} 
  V.~Khachatryan {\it et al.}  [CMS Collaboration],
  arXiv:1502.07400 [hep-ex].
  
  
\bibitem{Vicente:2015cka} 
  For a recent review see: A.~Vicente,
  arXiv:1503.08622 [hep-ph].
  
  
  
\end{thebibliography}
\end{document}